\documentstyle[12pt]{article}
\makeatletter
\def\@sect#1#2#3#4#5#6[#7]#8{\ifnum #2>\c@secnumdepth
    \def\@svsec{}\else
    \refstepcounter{#1}\edef\@svsec{\csname the#1\endcsname.\hskip 1em }\fi
    \@tempskipa #5\relax
    \ifdim \@tempskipa>\z@
    \begingroup #6\relax
    \@hangfrom{\hskip #3\relax\@svsec}{\interlinepenalty \@M #8\par}
    \endgroup
    \csname #1mark\endcsname{#7}\addcontentsline
    {toc}{#1}{\ifnum #2>\c@secnumdepth \else
     \protect\numberline{\csname the#1\endcsname}\fi
           #7}\else
    \def\@svsechd{#6\hskip #3\@svsec #8\csname #1mark\endcsname
          {#7}\addcontentsline
          {toc}{#1}{\ifnum #2>\c@secnumdepth \else
     \protect\numberline{\csname the#1\endcsname}\fi
           #7}}\fi
     \@xsect{#5}}
\def\label#1{\@bsphack\if@filesw {\let\thepage\relax
   \xdef\@gtempa{\write\@auxout{\string
   \newlabel{#1}{{\thesection.\@currentlabel}{\thepage}}}}}\@gtempa
   \if@nobreak \ifvmode\nobreak\fi\fi\fi\@esphack}
\def\@eqnnum{(\thesection.\theequation)}
\def\section{\setcounter{equation}{0} \@startsection {section}{1}{\z@}{-3.5ex
   plus -1ex minus -.2ex}{2.3ex plus .2ex}{\Large\bf}}
\newcount\@minsofar
\newcount\@min
\newcount\@cite@temp
\def\@citex[#1]#2{%
\if@filesw \immediate \write \@auxout {\string \citation {#2}}\fi
\@tempcntb\m@ne \let\@h@ld\relax \def\@citea{}%
\@min\m@ne%
\@cite{%
  \@for \@citeb:=#2\do {\@ifundefined {b@\@citeb}%
    {\@h@ld\@citea\@tempcntb\m@ne{\bf ?}%
    \@warning {Citation `\@citeb ' on page \thepage \space undefined}}%
{\@minsofar\z@ \@for \@scan@cites:=#2\do {%
  \@ifundefined{b@\@scan@cites}%
    {\@cite@temp\m@ne}
    {\@cite@temp\number\csname b@\@scan@cites \endcsname \relax}%
\ifnum\@cite@temp > \@min% select the next one to list
    \ifnum\@minsofar = \z@
      \@minsofar\number\@cite@temp
      \edef\@scan@copy{\@scan@cites}\else
    \ifnum\@cite@temp < \@minsofar
      \@minsofar\number\@cite@temp
      \edef\@scan@copy{\@scan@cites}\fi\fi\fi}\@tempcnta\@min
  \ifnum\@minsofar > \z@ % some more
    \advance\@tempcnta\@ne
    \@min\@minsofar
    \ifnum\@tempcnta=\@minsofar %   Number follows previous--hold on to it
      \ifx\@h@ld\relax
        \edef \@h@ld{\@citea\csname b@\@scan@copy\endcsname}%
    \else \edef\@h@ld{\ifmmode{-}\else--\fi\csname b@\@scan@copy\endcsname}%
      \fi
    \else \@h@ld\@citea\csname b@\@scan@copy\endcsname
          \let\@h@ld\relax
  \fi % no more
\fi}%
\def\@citea{,\penalty\@highpenalty\,}}\@h@ld}{#1}}
\def\appendixname{Appendix}
\def\appendix{\par
  \def\pre@section{\appendixname{} }
  \setcounter{section}{1}
  \@addtoreset{equation}{section}
  \def\thesection{\Alph{section}}
  \def\theequation{\arabic{equation}}}
\makeatother
\begin{document}
\def\t{\theta}
\def\T{\Theta}
\def\w{\omega}
\def\ov{\overline}
\def\a{\alpha}
\def\b{\beta}
\def\g{\gamma}
\def\s{\sigma}
\def\l{\lambda}
\def\wt{\widetilde}
\def\di{\displaystyle}
\def\sn{\mbox{sn}}
\def\cd{\mbox{cd}}
\def\cn{\mbox{cn}}
\def\dn{\mbox{dn}}

\addtolength{\unitlength}{-0.5\unitlength}

\newsavebox{\wfrontsq}\savebox{\wfrontsq}(75,75){\begin{picture}(75,75)
\thicklines\multiput(0,0)(0,75){2}{\line(1,0){75}}
\multiput(0,0)(75,0){2}{\line(0,1){75}}\end{picture}}

\newsavebox{\wleftsq}\savebox{\wleftsq}(37.5,112.5)
{\begin{picture}(37.5,112.5)
\thicklines\multiput(0,37.5)(0,75){2}{\line(1,-1){37.5}}
\multiput(0,37.5)(37.5,-37.5){2}{\line(0,1){75}}\end{picture}}

\newsavebox{\wupsq}\savebox{\wupsq}(112.5,37.5)
{\begin{picture}(112.5,37.5)
\thicklines\multiput(0,37.5)(75,0){2}{\line(1,-1){37.5}}
\multiput(0,37.5)(37.5,-37.5){2}{\line(1,0){75}}\end{picture}}

\newsavebox{\bfrontsq}\savebox{\bfrontsq}(75,75){\begin{picture}(75,75)
\thicklines\multiput(0,0)(0,75){2}{\line(1,0){75}}
\multiput(0,0)(75,0){2}{\line(0,1){75}}\put(0,0){\line(1,1){75}}
\multiput(0,50)(50,-50){2}{\line(1,1){25}}
\end{picture}}

\newsavebox{\bleftsq}\savebox{\bleftsq}(37.5,112.5)
{\begin{picture}(37.5,112.5)
\thicklines\multiput(0,37.5)(0,75){2}{\line(1,-1){37.5}}
\multiput(0,37.5)(37.5,-37.5){2}{\line(0,1){75}}
\put(37.5,0){\line(-1,3){37.5}}
\multiput(12.5,25)(25,25){2}{\line(-1,3){12.5}}
\end{picture}}

\newsavebox{\bupsq}\savebox{\bupsq}(112.5,37.5)
{\begin{picture}(112.5,37.5)
\thicklines\multiput(0,37.5)(75,0){2}{\line(1,-1){37.5}}
\multiput(0,37.5)(37.5,-37.5){2}{\line(1,0){75}}
\put(0,37.5){\line(3,-1){112.5}}
\multiput(62.5,0)(25,25){2}{\line(-3,1){37.5}}
\end{picture}}

\begin{flushright} October, 1993
\end{flushright}
\vspace{3cm}
\begin{center}
{\bf New series of 3D lattice integrable models}\\
\vspace{1cm}
V.V. Mangazeev, S.M. Sergeev, Yu.G. Stroganov\\
\vspace{1cm}
{\it Institute for High Energy Physics,}\\
{\it Protvino, Moscow Region, Russia}
\end{center}
\vspace{1cm}
\begin{flushleft}
{\bf Abstract}
\end{flushleft}
In this paper we present a new series of 3-dimensional integrable
lattice models with $N$ colors. The case $N=2$ generalizes
the elliptic model of paper \cite{Man}. The weight functions of the models
satisfy modified tetrahedron equations with $N$ states and give
a commuting family of two-layer transfer-matrices.
The dependence on the spectral parameters
corresponds to the static limit of the modified tetrahedron equations and
weights are parameterized in terms of elliptic functions.
The models contain two free parameters: elliptic modulus
and additional parameter $\eta$. Also we briefly discuss
symmetry properties of weight functions of the models.

\newpage

\section{Introduction.}

The purpose of this work is an enlarging  of still poor zoo of 3D integrable
lattice models. We say the model to be integrable if it possesses a family
of commuting transfer matrices.
The existence of such family is ensured, for example, by the construction of
solutions of the tetrahedron equation, which is a three dimensional
generalization of the Yang -- Baxter equation \cite{Z1,Z2,B,BS,JM}.
As an example we can mention the $N$ -- color trigonometric model by
Bazhanov -- Baxter \cite{BB}. Although its integrability has been proved
by a different method, the
Boltzmann weights of this model are the solutions of the tetrahedron equation
\cite{StSq}.
This solution, as well as the first known
Zamolodchikov's one \cite{Z1,Z2} (which is  a particular case of the
Bazhanov-Baxter model when $N=2$) can be parameterized in terms of
trigonometric functions depending on tetrahedron angles.
In our previous paper \cite{Man} we have constructed an elliptic
two -- color solution of a modified system of the tetrahedron equations,
which provides the commutativity of two -- layer transfer-matrices.
This solution in a sense is not ``full'' and corresponds to the case
of the so called ``static'' limit of tetrahedron equations.

In difference from Bazhanov -- Baxter model, for which
the solution of tetrahedron equations contains six angular variables
(five of them are independent), the static solution of the modified
tetrahedron equations from \cite{Man} can be parameterized by
three angle -- like variables and one additional parameter
(the modulus of elliptic functions).

In this paper we generalize this elliptic solution to the case of
an arbitrary number $N$ of spin values and obtain one more
parameter, on which weight functions depend.
This additional parameter is the same for all weights and new solutions
are still static. Boltzmann weights of the new models like
the weight functions of
Bazhanov -- Baxter model have the so called Body -- Centered -- Cube (BCC)
form, invented by Baxter in \cite{Bax}. This fact allows us to use
the technique of the Star -- Square relations, developed in \cite{StSq}.

The paper is organized as follows. In Section 2 we recall main definitions
and notations, give the form of the Boltzmann weights and write out the
modified tetrahedron equations.
In Section 3 we formulate conditions for the BCC ansatz for weight functions
to obey the modified tetrahedron equations.
In Section 4 a natural parameterization of the obtained
solution is given in terms of elliptic functions.
In Section 5 we discuss symmetry properties of weight functions of the model.
At last, Appendix contains a detailed consideration of $N=2$ case
and an explanation of the transition to the weights of the model from
paper \cite{Man}.

\section{Body-Centered-Cube (BCC) ansatz\hfill\break
         for weight functions}

In this section we recall some definitions from
\cite{KMS,StSq} and give the explicit form of weight functions.
We also write out modified tetrahedron equations, which provide
a commutativity of two-layer transfer matrices \cite{Man}.

Consider a simple cubic lattice ${\cal L}$ consisting of two types
of elementary cubes alternating in checkerboard order in all directions
(see Fig.~1).

\begin{picture}(700,385)
\put(50,55){
\begin{picture}(600,300)
\multiput(150,0)(150,0){3}{\usebox{\wfrontsq}}
\multiput(150,75)(150,0){3}{\usebox{\bfrontsq}}
\multiput(225,0)(150,0){3}{\usebox{\bfrontsq}}
\multiput(225,75)(150,0){3}{\usebox{\wfrontsq}}
\multiput(112.5,0)(-75,75){2}{\usebox{\wleftsq}}
\multiput(112.5,75)(-75,75){2}{\usebox{\bleftsq}}
\multiput(75,37.5)(-75,75){2}{\usebox{\bleftsq}}
\multiput(75,112.5)(-75,75){2}{\usebox{\wleftsq}}
\multiput(112.5,150)(150,0){3}{\usebox{\bupsq}}
\multiput(187.5,150)(150,0){3}{\usebox{\wupsq}}
\multiput(75,187.5)(150,0){3}{\usebox{\wupsq}}
\multiput(150,187.5)(150,0){3}{\usebox{\bupsq}}
\multiput(37.5,225)(150,0){3}{\usebox{\bupsq}}
\multiput(112.5,225)(150,0){3}{\usebox{\wupsq}}
\multiput(0,262.5)(150,0){3}{\usebox{\wupsq}}
\multiput(75,262.5)(150,0){3}{\usebox{\bupsq}}
\end{picture}}
\put(310,0){\Large\bf Fig. 1}
\end{picture}
\vspace{0.2cm}

At each site of ${\cal L}$ place a spin variable taking its values in $Z_N$,
for any integer $N\ge2$
(elements of $Z_N$ are given by $N$ distinct numbers $0,1,\ldots,N-1$
considered {\it modulo} $N$).
To each ``white'' cube we assign a weight
function $W(a|efg|bcd|h)$ (see Fig.~2)

\begin{picture}(600,265)
\put(0,50){
\begin{picture}(500,200)
\multiput(140,0)(120,0){2}{\line(0,1){120}}
\multiput(140,0)(0,120){2}{\line(1,0){120}}
\multiput(140,0)(0,120){2}{\line(-1,1){60}}
\put(80,180){\line(1,0){120}}\put(80,180){\line(0,-1){120}}
\put(200,180){\line(1,-1){60}}
\multiput(200,180)(0,-20){6}{\line(0,-1){12}}
\multiput(80,60)(20,0){6}{\line(1,0){12}}
\multiput(255,5)(-30,30){2}{\line(-1,1){20}}
\multiput(140,0)(120,0){2}{\circle*{15}}
\multiput(140,120)(120,0){2}{\circle*{15}}
\multiput(80,60)(120,0){2}{\circle*{15}}
\multiput(80,180)(120,0){2}{\circle*{15}}
\put(300,80){\large $=\quad W(a|e,f,g|b,c,d|h)$}
\put(150,10){$e$}\put(270,10){$d$}\put(150,100){$a$}
\put(270,100){$f$}\put(212,186){$b$}\put(92,190){$g$}
\put(92,65){$c$}\put(212,65){$h$}
\end{picture}
}
\put(320,0){\Large\bf Fig. 2}
\end{picture}
\vspace{0.2cm}

\noindent
and to each ``dashed'' cube -- a weight
function $\ov W(a|efg|bcd|h)$, where as usual $a,\ldots,h$ -- spin variables
placed in the vertices of each elementary cube.

Then the partition function of the model reads
\begin{equation}
Z=\sum_{spins}\prod_{"white"\atop cubes}W(a|efg|bcd|h)
\prod_{"dashed"\atop cubes}\ov W(a|efg|bcd|h).  \label{1}
\end{equation}

Following \cite{Man} suppose that weight functions $W$ and $\ov W$
satisfy the following equations:
\begin{eqnarray}
\sum_{d}
&W(a_4|c_2,c_1,c_3|b_1,b_3,b_2|d)\ov W'(c_1|b_2,a_3,b_1|c_4,d,c_6|b_4)&
\nonumber\\
\times&W''(b_1|d,c_4,c_3|a_2,b_3,b_4|c_5)
\ov W'''(d|b_2,b_4,b_3|c_5,c_2,c_6|a_1)&\nonumber\\
=\sum_{d}
&W'''(b_1|c_1,c_4,c_3|a_2,a_4,a_3|d)\ov W''(c_1|b_2,a_3,a_4|d,c_2,c_6|a_1)&
\nonumber\\
\times&W'(a_4|c_2,d,c_3|a_2,b_3,a_1|c_5)\ov W(d|a_1,a_3,a_2|c_4,c_5,c_6|b_4),&
                                                             \label{2}
\end{eqnarray}
where $W$, $W'$, $W''$, $W'''$ and $\ov W$, $\ov W'$, $\ov W''$, $\ov W'''$
are four independent pairs of weight functions. Suppose
that the dual variant of (\ref{2}) is also valid (with all $W$'s replaced
by $\ov W$'s and vice versa). We call this pair of equations
as a system of modified tetrahedron equations.
Note that if put $\ov W=W$ in (\ref{2}),
we come to the standard version of tetrahedron equations (Eqs. (2.2) of
\cite{B}). In the Bazhanov -- Baxter model all weights entering in
tetrahedron equations are parameterized in terms of six angles
$\t_1,\ldots,\t_6$ satisfying one quadrilateral constraint.
The explicit dependence on spectral parameters can be exhibited as
($\ov W=W$)
\begin{eqnarray}
&W\to W(\theta_2,\theta_1,\theta_3),\quad
W'\to W(\pi-\theta_6,\theta_1,\pi-\theta_4),&\nonumber\\
&W''\to W(\theta_5,\pi-\theta_3,\pi-\theta_4),\quad
W'''\to W(\theta_5,\theta_2,\theta_6).& \label{2a}
\end{eqnarray}
In the static limit three spectral parameters $\t_1,\t_2,\t_3$ in
weight functions $W(\t_1,\t_2,\t_3)$, $\ov W(\t_1,\t_2,\t_3)$ are
constrained by relation $\t_1+\t_2+\t_3=\pi$. Then formulas (\ref{2a}) are
transformed as follows:
\begin{eqnarray}
W\to W(\theta_2,\theta_1,\pi-\t_1-\t_2),\>
W'\to W(\t_2+\t_5,\theta_1,\pi-\t_1-\t_2-\t_5),&&\nonumber\\
W''\to W(\theta_5,\t_1+\t_2,\pi-\t_1-\t_2-\t_5),\>
W'''\to W(\theta_5,\theta_2,\pi-\t_2-\t_5)&& \label{2b}
\end{eqnarray}
and the same for $\ov W$ weights. Note that in this case the quadrilateral
constraint between tetrahedron angles is satisfied automatically.

As it is shown in \cite{Man} the system of modified tetrahedron equations
provides a commuting family of two-layer transfer-matrices.
Let us construct a horizontal transfer-matrix $T(W,\ov W)$ from alternating
weights $W$ and $\ov W$ as in Fig.~1. Here we imply that ingoing
indices of the transfer-matrix correspond to the spins of the lower layer,
outgoing indices -- to the spins of the upper one and the summation
over the spins of the middle layer is performed.
As usual partition function (\ref{1}) can be rewritten as
\begin{equation}
Z=Tr\{T(W,\ov W)]\}^M,\label{2c}
\end{equation}
where $2M$ is a number of horizontal layers of the lattice and we
imply periodic boundary conditions.
When equation (\ref{2}) and its dual variant are satisfied,
we have the following commutativity relation:
\begin{equation}
T(W,\ov W)T(W',\ov W')=T(W',\ov W')T(W,\ov W).               \label{3}
\end{equation}

Now let us specify the explicit form of weight functions. To make it,
denote
\begin{equation}
\w=\exp(2\pi i/N),\qquad \w^{1/2}=\exp(\pi i/N).                \label{4}
\end{equation}
Further, taking $x,y,z$ to be complex parameters constrained by the
Fermat equation
\begin{equation}
x^N+y^N=z^N                                                    \label{5}
\end{equation}
and $l$ to be an element of $Z_N$, define
\begin{equation}
w(x,y,z|l)=\prod_{s=1}^{l}{y\over z-x\w^s}.                    \label{6}
\end{equation}
In addition, define the function with one more argument
\begin{equation}
w(x,y,z|k,l)=w(x,y,z|k-l)\Phi(l),\quad k,l\in Z_N,               \label{7}
\end{equation}
where
\begin{equation}
\Phi(l)=\w^{l(l+N)/2}.                                          \label{8}
\end{equation}
Let us mention also two formulas for $w$ functions, which are useful for
calculations:
\begin{equation}
w(x,y,z|l+k)=w(x,y,z|k)w(x\w^k,y,z|l),                         \label{9}
\end{equation}
\begin{equation}
w(x,y,z|k,l)=\w^{kl}/w(z,\w^{1/2}y,\w x|l,k).                  \label{10}
\end{equation}

Now introduce the set of homogeneous variables $x_i$, $i=1,\ldots,8$ and
$x_{13}$, $x_{24}$, $x_{58}$, $x_{67}$ satisfying
\begin{equation}
x_{13}^N=x_1^N-x_3^N,\quad x_{24}^N=x_2^N-x_4^N,\quad
x_{58}^N=x_5^N-x_8^N,\quad x_{67}^N=x_6^N-x_7^N.                \label{11}
\end{equation}
Using all these notations we define the weight function $W(a|efg|bcd|h)$
as
\begin{equation}
W(a|e,f,g|b,c,d|h)=
\sum_{\sigma\in Z_N}{w(x_3,x_{13},x_1|d,h+\sigma)
w(x_4,x_{24},x_2|a,g+\sigma)
\over w(x_8,x_{58},x_5|e,c+\sigma)w(x_7/\w,x_{67},x_6|f,b+\sigma)}.
                                                                  \label{12}
\end{equation}

In fact, the Boltzmann weights of form (\ref{12}) generalize
the weight functions of the Bazhanov --Baxter model \cite{BB,KMS}.
Up to inessential gauge factors the latter corresponds to the choice
\begin{equation}
x_5=x_1,\quad x_6=x_2,\quad x_7=x_3, \quad x_8=x_4. \label{12a}
\end{equation}
Following paper \cite{BB} we will call formula (\ref{12}) as
a Body-Centered-Cube (BCC) ansatz for weight functions.

\section{Star-square relation and the proof of modified
tetrahedron equations}

In paper \cite{StSq} tetrahedron equations for the Bazhanov -- Baxter
model with $N$ states were proved using the so called Star-Square and
``inversion'' relations. We will follow  the method of this paper
for weight function (\ref{12}).

First recall ``inversion'' relation for functions $w(x,y,z|k,l)$:
\begin{equation}
\sum_{k\in Z_N}{w(x,y,z|k,l)\over w(x,y,\w z|k,m)}=
N\delta_{l,m}{(1-z/x)\over (1-z^N/x^N)},                    \label{13}
\end{equation}
where $l,\>m\in Z_N$, $x,y,z$ satisfy (\ref{5}) and $\delta_{l,m}$ is
the Kronecker symbol on $Z_N$.

To write down the Star-Square relation introduce
a non-cyclic analog of $w$ function, defined recurrently as follows:
\begin{equation}
w(x|0)=1,\quad{w(x|l)\over w(x|l-1)}={1\over(1-x\w^l)},\quad l\in Z,
                                                            \label{14}
\end{equation}
where $Z$ is the set of all integers.
It is obvious that
\begin{equation}
w(x,y,z|l)=
{\Bigl(}{y\over z}{\Bigr)}^l w(x/z|l),\quad l\in Z_N, \label{15}
\end{equation}
where index $l$, being considered modulo $N$, is interpreted as an element
of $Z_N$.
Then we have the following identity
\begin{eqnarray}
&\biggl\{{\di
\sum_{\s\in Z_N} {{w(x_1,y_1,z_1|a+\s)w(x_2,y_2,z_2|b+\s)}\over
{w(x_3,y_3,z_3|c+\s)w(x_4,y_4,z_4|d+\s)}}}\biggr\}_0=&\nonumber\\
&{\di
{(x_2z_1/x_1z_2)^{a-b}(x_1y_2/x_2z_1)^b(z_3/y_3)^c(z_4/y_4)^d\over
\Phi (a-b)\w^{(a+b)/2}}}&\nonumber\\
&\times{\di{w(\w x_3x_4z_1z_2/x_1x_2z_3z_4|c+d-a-b)}\over
\di{w({x_4z_1\over x_1z_4}|d-a)w({x_3z_2\over x_2z_3}|c-b)
w({x_3z_1\over x_1z_3}|c-a)w({x_4z_2\over x_2z_4}|d-b)}}&,
                                                               \label{16}
\end{eqnarray}
where the lower index ``$0$'' after the curly brackets indicates that the
l.h.s. of (\ref{16}) is normalized to unity at zero exterior spins, and
the following constraint is imposed
\begin{equation}
{y_1\over z_1}{y_2\over z_2}{z_3\over y_3}{z_4\over y_4}=\w.    \label{17}
\end{equation}
We call relation (\ref{16}) as the Star-Square one.
Note that the separate $w$'s in the r.h.s. of (\ref{16})
are not single-valued functions on $Z_N$, while the whole
expression is cyclic on the exterior spins $a,b,c,d$.

The proof of (\ref{13}), (\ref{16}) was given in \cite{StSq} and we refer
the interested reader to this paper.

Now we turn to relation (\ref{2}). Instead of
weights $W$, $W'$, $W''$, $W'''$ let us substitute into (\ref{2})
explicit formula (\ref{12}) with corresponding sets of parameters:
\begin{eqnarray}
W,W',W'',W'''\to W(x_i,x_{ij}), W(x'_i,x'_{ij}),
W(x''_i,x''_{ij}), W(x'''_i,x'''_{ij}),&&\nonumber\\
\ov W,\ov W',\ov W'',\ov W'''\to W(\ov x_i,\ov x_{ij}),
W(\ov x'_i,\ov x'_{ij}),
W(\ov x''_i,\ov x''_{ij}),W(\ov x'''_i,\ov x'''_{ij}).&&\label{18}
\end{eqnarray}

Let us multiply both sides of (\ref{2}) by the following product of $w$
weights:
\begin{eqnarray}
&{\di w(x''_7,x''_{67},x''_6|c_4,a_2+l_2)\over\di
w(\ov x'''_3/\w,\ov x'''_{13},
\ov x'''_1|c_6,a_1+l_1)}{\di
w(\ov x'_7,\w\ov x'_{67},\ov x'_6|a_3,c_4+l_3)\over
\di w(x_4,x_{24},\w x_2|a_4,c_3+l_4)}&\times\nonumber\\
\times&{\di w(\ov x''_8,\ov x''_{58},\ov x''_5/\w|b_2,c_2+m_2)\over
\di w(x'''_4,
x'''_{24},\w x'''_2|b_1,c_3+m_1)}{\di w(\w x'_8,x'_{58},x'_5|c_2,b_3+m_3)\over
\di w(\ov x_3/\w,\ov x_{13},\ov x_1|c_6,b_4+m_4)}&                  \label{19}
\end{eqnarray}
and sum over $a_1,a_2,a_3,a_4,\>b_1,b_2,b_3,b_4$.
Note that due to  ``inversion'' relation (\ref{13}) we do not
lose any information after such transition from spins
$a_i,b_i$ to $l_i,m_i$.

The functions $w$ in expression (\ref{19}) are chosen in such a way
that using relation (\ref{13}) we can calculate the sums over the spins
$a_1,a_2,a_3,a_4$ in the l.h.s. and those over $b_1,b_2,b_3,b_4$ in the
r.h.s. of the obtained equation and cancel the summations over the
spins $\sigma$'s, which come from expression (\ref{12})
for the functions $W$'s and $\ov W$'s.
Now let us consider applicability conditions (\ref{17}) of
Star-Square relation (\ref{16}) for the sums over $a_1,a_2,a_3,a_4$
in the r.h.s. and over $b_1,b_2,b_3,b_4$ in the l.h.s.
of the obtained equation.
We obtain eight conditions on $x$'s and $\ov x$'s.
Applying relation (\ref{16}) eight times and calculating the sums
over $d$ spin in the l.h.s. and r.h.s (the spin structure of the sums
over $d$ has the form of relation (\ref{13}) and we demand that
corresponding variables $x,y,z$ entering in the arguments of functions
$w(x,y,z|k,l)$ are constrained in  such a way that relation
(\ref{13}) can be used), we come to the equation without any summation.
The l.h.s. and r.h.s. of this equation consist of the products of ten $w$
functions  with the same spin structure and
expressions like $x_1^{m_2}$ coming from relation (\ref{16}).
Let us impose all necessary constraints on parameters $x_i$, $x_{ij}$,
$\ov x_i$ and $\ov x_{ij}$
to satisfy this equation.
On obeying of these constraints one can show that
spin independent multipliers coming from relations (\ref{13}), (\ref{16})
coincide.

We also have to satisfy the dual variant of (\ref{2}). Hence, we must
add a dual set with $x$ replaced by $\ov x$ and vice versa
to all obtained constraints on parameters $x$ and $\ov x$ .
We will not write out here all these relations.
A detailed analysis shows that we have two solutions.
The first one corresponds to the choice
\begin{equation}
\ov x_i=x_i,\>\ov x_{13}=x_{13},\>\ov x_{24}=x_{24},\>\ov x_{58}=x_{58},\>
\ov x_{67}=x_{67},\>i=1,\ldots,8                         \label{20}
\end{equation}
and the same for $x'$, $x''$, $x'''$. This choice corresponds to the
Bazhanov -- Baxter model, considered in \cite{StSq}.

But there is also another possibility. It will be convenient to fix
a normalization of all parameters $x$ in $W$ functions as
\begin{equation}
x_3=1,\quad x_4=1,\quad x_7=1,\quad x_8=1                 \label{21}
\end{equation}
for all sets $x$, $x'$, $x''$, $x'''$ and
$\ov x$, $\ov x'$, $\ov x''$, $\ov x'''$.
Then all parameters $\ov x$ can be expressed in terms of $x$ as:
\begin{eqnarray}
&\ov x_1=1/x_2,\quad \ov x_2=1/x_1,\quad \ov x_5=1/x_6,\quad
\ov x_6=1/x_5,&\nonumber\\
&\ov x_{13}=\w^{-1/2}{\di x_{24}\over\di x_2},
\> \ov x_{24}=\w^{1/2}{\di x_{13}\over\di x_1},\>
\ov x_{58}=\w^{1/2}{\di x_{67}\over\di x_6},\>
\ov x_{67}=\w^{-1/2}{\di x_{58}\over\di x_5}.& \label{22}
\end{eqnarray}
Let us introduce the following notations:
\begin{eqnarray}
&s={\di{x_1x_2}\over\di{x_5x_6}},\quad t_1={\di x_5\over{\di\w x_1}},
\quad t_2={\di x_6\over\di x_1},
\quad t_3={\di{x_{58}x_{67}}\over\di{x_{13}x_{24}}},&\nonumber\\
&j_1={\di x_6\over\di x_1}{\di{x_{13}x_{58}}\over\di{x_{67}x_{24}}},\quad
j_2=\w{\di x_5\over\di x_1}{\di{x_{13}x_{67}}\over\di{x_{58}x_{24}}},\quad
j_3=x_5x_6.&                                                     \label{23}
\end{eqnarray}
and the same for the sets of $x'$, $x''$, $x''$.
Then all constraints on parameters take the form
\begin{eqnarray}
&t_1=t'''_2,\quad t_2=t'_2,\quad t_3=t''_2,\quad t'_1=t'''_3,\quad
t'_3=t''_3,\quad t''_1=t'''_1,&\nonumber\\
&j_1=j'''_2,\quad j_2=j'_2,\quad j_3=j''_2,\quad j'_1=j'''_3,\quad
j'_3=j''_3,\quad j''_1=j'''_1,&\label{24}\\
&s=s'=s''=s''',\quad x_{24}x''_{24}=x'_{24}x'''_{24}.\nonumber
\end{eqnarray}
Relations (\ref{21}-\ref{24}) together with consistency relations (\ref{11})
are sufficient conditions for weight functions
(\ref{12}) to satisfy modified tetrahedron equations (\ref{2}) and
their dual variant. In the next section we will obtain a natural
parameterization for relations (\ref{11}), (\ref{23}-\ref{24}) in terms
of elliptic functions.

\section{Parameterization}

Recall that consistency relations (\ref{11}) connect $N$-th powers of
$x_i$ and $x_{ij}$. Hence, it is convenient to consider $N$-th powers
of (\ref{23}-\ref{24}) and introduce new parameters
\begin{equation}
S^2=s^N,\quad T^2_i=-t_i^N/S,\quad J^2_i=j_i^N,\quad i=1,\ldots,3. \label{25}
\end{equation}
{}From (\ref{23}-\ref{24}) we see that  parameter $S$ is the same for
all weights entering in relation (\ref{2}).

Parameters $S$, $T_i$, $J_i$, $i=1,2,3$  are defined by four independent
variables $x_1$, $x_2$, $x_5$, $x_6$. Therefore, there are three
additional constraints between these values. After simple calculations
we obtain
\begin{equation}
J_3={\di T_1J_1-T_2J_2\over\di T_1J_2-T_2J_1},\quad
T_3={\di (T_1^2-T_2^2)(1+SJ_1J_2T_1T_2)\over\di(1-T_1^2T_2^2)
(T_1J_2-T_2J_1)},   \label{26}
\end{equation}
and
\begin{equation}
P\equiv{\di S+T_i^2-SJ_i^2(1+ST_i^2)\over T_iJ_i},
\quad i=1,2,\label{27}
\end{equation}
where we introduce a new parameter $P$.

Note that definitions (\ref{25}) contain only the second powers of
variables $S$, $T_i$, $J_i$ and in (\ref{26}-\ref{27})
we have chosen some signs in a convenient way.
Using (\ref{26}) it is easy to obtain that we can add the case $i=3$
to relations (\ref{27}). The validity of formulas (\ref{26}-\ref{27})
can be checked by substitution of relations (\ref{23}), (\ref{25}) and
consistency conditions (\ref{11}).

Introduce an elliptic curve
\begin{equation}
P={\di S+x^2-Sy^2(1+Sx^2)\over \di xy}.                 \label{28}
\end{equation}
Points $(T_i,J_i)$, $i=1,2,3$ belong to this curve. We can
uniform it in terms of elliptic functions (see, for example, \cite{Baxbook}).
Note that formulas (\ref{26}) take the form of addition theorems
for elliptic functions.
Then we obtain
\begin{eqnarray}
&T_j=im^{1/2}\sn(u_j),\quad J_j={\di\sn(\eta -u_j)\over\di\sn(\eta)},\quad
\quad j=1,\ldots,3,&\nonumber\\
&P=-2im^{1/2}\sn(\eta)\cn(\eta)\dn(\eta),\quad S=m\,\sn^2(\eta).&  \label{29}
\end{eqnarray}
where $\sn$, $\cn$, $\dn$ are elliptic functions of modulus $m$ and
\begin{equation}
u_3=u_1+u_2. \label{30}
\end{equation}
In such a way we can obtain the parameterization for $x_i^N$ and $x_{ij}^N$.
When we take roots of $N$-th powers, a phase ambiguity appears.
Making an appropriate choice of these phases so that (\ref{23}-\ref{24})
to  valid we obtain the following formulas
\begin{eqnarray}
&x_1=\w^{-1/2}\biggl\{{\di\sn(\eta-u_3)\sn(\eta)\over\di\sn(u_1)\sn(u_2)}
\biggr\}^{1/N},&\nonumber\\
&x_2=\w^{1/2}\bigl\{m^2\sn(\eta-u_3)\sn(\eta)\sn(u_1)\sn(u_2)\bigr\}^{1/N},
&\nonumber\\
&x_5=\w^{1/2}\biggl\{{\di\sn(u_1)\sn(\eta-u_3)\over\di\sn(u_2)\sn(\eta)}
\biggr\}^{1/N},\quad
x_6=\w^{-1/2}\biggl\{{\di\sn(u_2)\sn(\eta-u_3)\over\di\sn(u_1)\sn(\eta)}
\biggr\}^{1/N},&\nonumber\\\label{32}
\end{eqnarray}
and
\begin{eqnarray}
&x_{13}=\w^{-1/2}\biggl\{{\di\T(u_3)\T(0)H(\eta-u_1)H(\eta-u_2)\over
\di\T(\eta-u_3)\T(\eta)H(u_1)H(u_2)}\biggr\}^{1/N},&\nonumber\\
&x_{24}=\w^{1/2}\bigl\{{\di\T(u_3)\T(0)\T(\eta-u_1)\T(\eta-u_2)\over
\di\T(\eta-u_3)\T(\eta)\T(u_1)\T(u_2)}\bigr\}^{1/N},&\nonumber\\
&x_{58}=\w^{1/2}\biggl\{{\di H(u_3)\T(0)H(\eta-u_1)\T(\eta-u_2)\over
\di\T(\eta-u_3)H(\eta)\T(u_1)H(u_2)}\biggr\}^{1/N},&\nonumber\\
&x_{67}=\w^{-1/2}\biggl\{{\di H(u_3)\T(0)\T(\eta-u_1)H(\eta-u_2)\over
\di\T(\eta-u_3)H(\eta)H(u_1)\T(u_2)}\biggr\}^{1/N},&\label{33}
\end{eqnarray}
where $H(u)$ and $\T(u)$ are Jacobi elliptic functions (see \cite{Baxbook}).
Suppose that real parameters $u_1$, $u_2$, $\eta$ satisfy
\begin{equation}
0 < u_{1,2,3} <\eta<2{\cal M},\label{31}
\end{equation}
where ${\cal M}$ is the complete integral of the first kind of the
modulus $m$. These conditions guarantee that
all values of the elliptic functions in
(\ref{32}-\ref{33}) are non-negative, and we choose the positive values
of all roots of $N$-th power in formulas (\ref{32}-\ref{33}).

Now we can rewrite the Boltzmann weight as a function of four new
parameters: $W(u_1,u_2|m,\eta)$ (we omit the dependence from spin
variables). It is on easy to check that dual weights
$\ov W$ can be obtained by the shift of parameter $\eta$:
\begin{equation}
\ov W(u_1,u_2|m,\eta)=W(u_1,u_2|m,\eta+i{\cal M}'),     \label{34}
\end{equation}
where ${\cal M}'$ is the complete integral of the first kind of the
complementary modulus $m'=\sqrt{1-m^2}$. Equations (\ref{24}) relate
the sets of new parameters in different weights. Then we obtain
\begin{eqnarray}
&W\to W(u_2,u_1),\quad W'\to W(u_2+u_5,u_1),&\nonumber\\
&W''\to W(u_5,u_1+u_2),\quad W'''\to W(u_5,u_2).&           \label{35}
\end{eqnarray}
Parameters $m$ and $\eta$ are the same for all weight functions
and we omit them in (\ref{35}). Also note that we have chosen three
independent parameters in (\ref{35}) as $u_1$, $u_2$ and $u_5$
to emphasize the connection of our parameterization with static limit
one (\ref{2b}).

\section{Symmetry Properties}

In this section we discuss symmetry properties of weight functions (\ref{12})
of the model with respect to the group $G$ of transformations of a
three-dimensional cube (see \cite{KMS,BB2}). Rcall briefly some definitions.
Group $G$ has two generating elements $\rho$ and $\tau$.
Any element $\alpha$ of $G$ can be expressed as a composition of these
two elements. We define the action of elements $\tau$ and $\rho$ on
the set of spins $\{a|e,f,g|b,c,d|h\}$ as follows
\begin{equation}
\tau\{a|e,f,g|b,c,d|h\}=\{a|f,e,g|c,b,d|h\} \label{36}
\end{equation}
and
\begin{equation}
\rho\{a|e,f,g|b,c,d|h\}=\{g|c,a,b|f,h,e,|d\}. \label{36a}
\end{equation}
Further it will be convenient to remove normalization conditions (\ref{21})
and to restore the homogeneity of the parameters $x$'s.

It is easy to see that weight function (\ref{12}) is invariant under
the action of element $\tau$, if we make the following transformation
of parameters:
\begin{eqnarray}
&(x_3,x_{13},x_1)\to(x_3,x_{13},x_1),\quad(x_4,x_{24},x_2)\to
(x_4,x_{24},x_2),&\nonumber\\
&(x_8,x_{58},x_5)\to(x_7/\w,x_{67},x_6),\quad(x_7,x_{67},x_6)\to(\w x_8,
x_{58},x_5).&  \label{37}
\end{eqnarray}
The action of element $\rho$ is less trivial and can be obtained with
the help of the Fourier transformation (see \cite{KMS}).
Introduce two more additional parameters $u$ and $v$ which are defined as
\begin{equation}
u^N={\di x_3^Nx_5^N-x_1^Nx_7^N\over\di x_1^N},\quad
v^N={\di x_2^Nx_8^N-x_4^Nx_6^N\over\di x_2^N}. \label{38}
\end{equation}
A phases of parameters $u$ and $v$ can be chosen in an arbitrary way.
Then one can show that the following identity is valid:
\begin{eqnarray}
&\Bigl\{\di\sum_{\sigma\in Z_N}\di{w(x_3,x_{13},x_1|d,h+\sigma)
w(x_4,x_{24},x_2|a,g+\sigma)
\over\di w(x_8,x_{58},x_5|e,c+\sigma)w(x_7/\w,x_{67},x_6|
f,b+\sigma)}\Bigr\}_0=&\nonumber\\
&=\w^{dh+ch-bf-bg}{\di w(x_4x_6,x_2v,x_2,x_8|a+b,f+g)\over
\di w(x_1x_7,x_1u,x_3x_5|e+h,d+c)}\times&\label{39}\\
&\times\Bigl\{\di\sum_{\sigma\in Z_N}
\di{w(x_7x_{13},x_1u,x_3x_{57}|e,d+\sigma)
w(x_2x_{68},x_2v,x_6x_{24}|g,b+\sigma)
\over\di w(x_5x_{13},\w x_1u,\w x_1x_{57}|c,h+\sigma)
w(x_4x_{68},x_2v,x_8x_{24}|a,f+\sigma)}\Bigr\}_0,&  \nonumber
\end{eqnarray}
where the lower index "$0$" after the curly brackets implies
that the expression in the curly brackets is divided by itself with all
exterior spin variables equated to zero.

The arrangement of spins in the sum of the r.h.s. in (\ref{39}) corresponds
to the $\rho$-transformed set of the indices (see (\ref{36a})). Hence, we
can define the action of $\rho$ transformation on the parameters as
\begin{eqnarray}
&(x_3,x_{13},x_1)\stackrel{\rho}{\rightarrow}(x_7x_{13},x_1u,x_3x_{57}),
\quad
(x_4,x_{24},x_2)\stackrel{\rho}{\rightarrow}(x_2x_{68},x_2v,x_6x_{24}),&
\nonumber\\
&(x_7,x_{57},x_5)\stackrel{\rho}{\rightarrow}(x_5x_{13},\w x_1u,\w x_1x_{57}),
&\nonumber\\
&(x_8,x_{68},x_6)\stackrel{\rho}{\rightarrow}(\w x_4x_{68},x_2v,x_8x_{24}).&
\label{40}
\end{eqnarray}
Using (\ref{39}) we can choose gauge factors before formula (\ref{12})
in such a way that the whole $W$ function will be invariant
under the transformations from the group $G$.
We will not write this symmetric form of weight
functions here. Note only that after $\rho$ transformation (\ref{40})
$W$ function is transformed into $\ov W$ and vice versa.
Using relations (\ref{32}-\ref{33}) from the previous section
we can write the following transformation rules for parameters $u_1,u_2,\eta$
with respect to the elements $\tau$ and $\rho$ from the group $G$:
\begin{equation}
u_1\stackrel{\tau}\rightarrow{u_2},\quad
u_2\stackrel{\tau}\rightarrow{u_1},\quad
\eta\stackrel{\tau}\rightarrow{\eta},\label{41}
\end{equation}
\begin{equation}
u_1\stackrel{\rho}\rightarrow{i{\cal M}'-u_1},\quad
u_2\stackrel{\rho}\rightarrow{u_1+u_2-i{\cal M}'},\quad
\eta\stackrel{\rho}\rightarrow{\eta+i{\cal M}'}.\label{42}
\end{equation}
We hope to use these symmetries for the calculation of the partition
function for our models.

\appendix

\section*{Appendix. A detailed consideration for the case $N=2$.}

The BCC form of the weight is useful for proving the modified
tetrahedron equations. But for other purposes a symmetrical form of the
Boltzmann weights, which differs from the BCC one
by some gauge transformation,
is more preferable. In this section we find this gauge multipliers
for the case when $N=2$ and obtain a table of Boltzmann weights
analogous to Baxter's table from \cite{Bax} and  endowed by obvious
symmetrical properties. Also we  show that fixing $\eta = {\cal M}$
we obtain the particular case considered in \cite{Man}.

We start from a simple gauge transformation of our BCC weight (\ref{12})
and consider the weight $W'$:
\begin{eqnarray}
&{\di W'(a|e,f,g|b,c,d|h)=
(-1)^{bf+ec+a+h}i^{-af+ch-dh+ag+cg-df}}\times&\nonumber\\
&\times{\di{Y_{e,d,h,c}Y_{g,c,e,a}Y_{c,h,b,g}
\over Y_{a,f,b,g}Y_{b,h,d,f}Y_{e,d,f,a}}
W(a|e,f,g|b,c,d|h)
},&\label{A1}
\end{eqnarray}
where
\begin{equation}
{\di Y_{a,f,b,g}=\exp\{-i{\pi\over 2}fg\}\exp\{i\pi ab(f-g)\}}\label{A2}
\end{equation}
It is convenient to introduce spins $(-1)^a =\pm 1$
instead of $a=0,1$. Further we shall use only the
multiplicative spins and write $a$ instead of $(-1)^a$, etc.

Absolute values of these weights ( both $W$ and $W'$)
depend only on three Baxter's sign variables $abeh$, $acfh$, $adgh$, $abcd$.
Therefore up to signs there exist sixteen different weights (\ref{A1}).
In terms of multiplicative spins we can write the
following parameterization:
\begin{eqnarray}
&W'=\xi W_{S}\times&\nonumber\\
\times&\exp\{(M_1+N_1)agbf+(M_1-N_1)cedh+&\nonumber\\
&+(M_2+N_2)agce+(M_2-N_2)bfdh+&\nonumber\\
&+(M_3+N_3)aefg+(M_3-N_3)cgbf+&\nonumber\\
&+(M_0+N_0)abcd +(M_0-N_0)efgh\},&\label{A4}
\end{eqnarray}
where, calculating all sign factors, we parameterize $W_{S}$
as follows:

\vskip 0.5cm

% table
\begin{center}
\begin{tabular}{||c|c|c||l||}
\hline
&&&\\
$abeh$&$acfh$&$adgh$&$ W_{S}(a|efg|bcd|h)$\\ &&&\\ \hline
$+$&$+$&$+$&$P_0$\\
$-$&$+$&$+$&$R_1$\\
$+$&$-$&$+$&$R_2$\\
$+$&$+$&$-$&$R_3$\\
$+$&$-$&$-$&$abP_1$\\
$-$&$+$&$-$&$acP_2$\\
$-$&$-$&$+$&$adP_3$\\
$-$&$-$&$-$&$abcdS_0$\\ \hline
\end{tabular}
\end{center}
\vskip 0.5cm

Parameter $\xi$ in (\ref{A4}) is a normalization factor and we choose it
so that $P_0=1$. Now we can obtain $\exp(16M_i)$,
$\exp(16N_i)$ and $W_{S}^2$ just by multiplying and dividing BCC weights for
several sets of the spins.

It is useful to express the answer in terms of elliptic functions of
another modulus $k$ instead of the used functions of the modulus $m$
\begin{equation}
k={1-m\over 1+m}.
\end{equation}
Denoting the complete elliptic integrals of the first kind of the moduli
$m$ and $m'=\sqrt{1-m^2}$ as $\cal M$ and ${\cal M}'$, and of the moduli
$k$  and $k'=\sqrt{1-k^2}$ as $\cal K$ and
${\cal K}'$, we have the connection between them
\begin{equation}
{\cal K}' = (1+m){\cal M},\;\; 2{\cal K} = (1+m){\cal M}'.
\end{equation}
One can easily obtain the reparameterization of $x_i^2$ and $x_{ij}^2$
using
\begin{equation}
{\di im^{1/2}\sn(v,m) = k'{\sn(x,k)\over\cn(x,k)\dn(x,k)},
\;\;\mbox{where}\;\; {2x\over 1+m}=iv.}\label{A5}
\end{equation}
To calculate expressions for $W_S$, $\exp(2M_i)$ and $\exp(2N_i)$,
we need three more complicated identities:
\begin{eqnarray}
&{\di {\sn(v_1+v_2)-\sn(v_1)-\sn(v_2)-m\sn(v_1)\sn(v_2)\sn(v_1+v_2)
\over\sn(v_1+v_2)+\sn(v_1)+\sn(v_2)-m\sn(v_1)\sn(v_2)\sn(v_1+v_2)}}&
\nonumber\\
&{\di={\sn(x_1)\sn(x_2)\over\cd(x_1)\cd(x_2)},}&\nonumber\\
&{\di {\sn(v_1)+\sn(v_2)+\sn(v_1+v_2)+m\sn(v_1)\sn(v_2)\sn(v_1+v_2)
\over\sn(v_1)+\sn(v_2)+\sn(v_1+v_2)-m\sn(v_1)\sn(v_2)\sn(v_1+v_2)}}&
\nonumber\\
&{\di={\cd(x_1+x_2)\over\cd(x_1)\cd(x_2)},}&\nonumber\\
&{\di {\sn(v_1)+\sn(v_2)-\sn(v_1+v_2)-m\sn(v_1)\sn(v_2)\sn(v_1+v_2)
\over\sn(v_1)+\sn(v_2)+\sn(v_1+v_2)-m\sn(v_1)\sn(v_2)\sn(v_1+v_2)}=}&
\nonumber\\
&{\di =-k^2\sn(x_1)\sn(x_2)\cd(x_1+x_2)},&\nonumber\\
&&\label{A6}
\end{eqnarray}
where in the LHS we imply the modulus $m$, and in the RHS -- $k$,
the arguments $v_i$ and $x_i$ being connected by (\ref{A5}).

We would like to obtain a model with real weights $W_S$. To do this we have
to consider a regime when $u_1$, $u_2$, $u_3$ are pure imaginary satisfying
\hbox{$0<\mbox{Im}u_{1,2,3}<{\cal M}'$}
and $\eta = {\cal M}+i\epsilon$, a small
$\epsilon$ being real.
Defining further the variables $z_{1,2,3},\lambda$ by
\begin{equation}
{2z_{1,2}\over 1+m} = -iu_{1,2},\; z_1+z_2+z_3 = K,\;
{2\lambda\over 1+m} = i\eta;\label{A7}
\end{equation}
and using repeatedly (\ref{A6}), we obtain the exponentials
\begin{eqnarray}
&\exp{2N_1}=e^{i\pi/4}\{-ik\sn(\lambda +z_1)\cd(\lambda +z_1)\}^{1/4},&
\nonumber\\
&\exp{2N_2}=\{-ik\sn(\lambda +z_2)\cd(\lambda +z_2)\}^{1/4},&\nonumber\\
&\exp{2N_3}=\{-ik\sn(\lambda -z_3)\cd(\lambda -z_3)\}^{1/4},&\nonumber\\
&\exp{2N_0}=\{-ik\sn(\lambda)\cd(\lambda)\}^{1/4};&\label{A8}\\
&{\di \exp{2M_1}=\{i{\cd(\lambda +z_1)\over\sn(\lambda +z_1)}\}^{1/4},\;\;
\exp{2M_2}=\{i{\cd(\lambda +z_2)\over\sn(\lambda +z_2)}\}^{1/4},}&\nonumber\\
&{\di \exp{2M_3}=\{i{\cd(\lambda -z_3)\over\sn(\lambda -z_3)}\}^{1/4},\;\;
\exp{2M_0}=\{-i{\sn(\lambda)\over\cd(\lambda)}\}^{1/4}.}&\label{A9}
\end{eqnarray}
In this formulae $z_i$ are real, $0<z_i<K$, and $\lambda = i{\cal K}'/2
+\epsilon'$, $\epsilon'$ being small and real. The expressions contained
in the curly brackets belong to the first or fourth quadrants of the complex
plane, the fourth power roots of these values are defined so that they belong
to the same quadrants.
As to $W_{S}$, it coincides with the elliptic solution of ref. \cite{Man}:
\begin{eqnarray}
&{\di P_0=1,\;\; P_i=\sqrt{\sn(z_j)\sn(z_k)\over\cd(z_j)\cd(z_k)},}&
\nonumber\\
&{\di S_0=k\sqrt{\sn(z_1)\sn(z_2)\sn(z_3)},\; \;
R_i=\sqrt{\sn(z_i)\over\cd(z_j)\cd(z_k)},}&
\label{A10}
\end{eqnarray}
where $i,j,k$ are any permutation of $1,2,3$. Here the square roots of
the positive values are supposed to be also positive.
Our weights coincide with those of \cite{Man} when
\begin{equation}
\lambda = i{{\cal K}'\over 2}\Leftrightarrow \eta = {\cal M}.\label{A11}
\end{equation}

The dual weights $\overline W$ can be obtained from these ones by the changing
$M_i\rightarrow -M_i$, $i=0,...,3$,
and $S_0\rightarrow -S_0$, so that the terms
$M_{1,2,3}$ and $N_{1,2,3}$ are the pure gauge of the system of the modified
tetrahedron equations and the only difference with the case (\ref{A11})
consists in the $M_0$ and $N_0$ terms.

Note that the extraction of the $\lambda$ -- dependent terms $\exp(M_0)$
and $\exp(N_0)$ as common multipliers of the $\lambda$ -- independent
weight $W_{S}$ is the property of the choice $N=2$, when $N\neq 2$ this
property fails.

\end{document}